\documentclass[prb,aps,twocolumn,amsmath,amssymb,floatfix,showpacs,
superscriptaddress]{revtex4}

\usepackage[dvips]{graphics}
\usepackage{color}
\definecolor{dred}{rgb}{0,0,0.6}

\begin{document}

\title{\textcolor{dred}{Verification of cosine squared relation of
electronic conductance in a biphenyl molecule}}

\author{Santanu K. Maiti}

\email{santanu.maiti@isical.ac.in}

\affiliation{Physics and Applied Mathematics Unit, Indian Statistical
Institute, 203 Barrackpore Trunk Road, Kolkata-700 108, India}

\begin{abstract}

The experimentally obtained (Venkataraman {\em et al}.~\cite{latha}) cosine 
squared relation of electronic conductance in a biphenyl molecule is 
verified theoretically within a tight-binding framework. Using Green's 
function formalism we numerically calculate two-terminal conductance as a
function of relative twist angle among the molecular rings and find that
the results are in good agreement with the experimental observation.

\end{abstract}


\maketitle

In a glorious experiment Venkataraman {\em et al.}~\cite{latha} have 
established that electronic conductance of a molecular wire does not
depend only on the chemical properties of the molecule used, but
also on its conformation. It has been examined that for the biphenyl 
molecule where two benzene rings are connected by a single C-C bond, 
electronic conductance varies significantly with the relative twist 
angle among these molecular rings. The conductance reaches to a maximum
for the planar conformation, while it gets reduced with increasing the 
twist angle and eventually drops to zero when the molecular rings
are perpendicular to each other. The experimental results suggest a clear 
correlation between junction conductance and molecular conformation which 
predicts that the conductance of the biphenyl molecule decreases with 
increasing twist angle obeying a {\em cosine squared relation}.

In this present communication we essentially verify theoretically this 
conformation dependent molecular conductance and prove that our numerical 
results agree well with the experimental realization. A simple tight-binding 
(TB) Hamiltonian is given to describe the model quantum system and we 
numerically compute molecular conductance using Green's function approach 
based on the Landauer conductance formula~\cite{land}. Within a 
non-interacting electron picture this framework is well applicable for
analyzing electron transport through a molecular bridge system, as 
illustrated by Aviram and Ratner~\cite{aviram} in their work where they 
have first described two-terminal electron transport through a molecule 
coupled to two metallic electrodes. Following this pioneering work later
many theoretical~\cite{nitzan1,woi,tagami,orella1,orella2,arai,walc,san1,
san2,san3,san4,san5,mag,lau,baer1,baer2,baer3,gold} as well as 
experimental~\cite{reed2,reed1,tali,fish,cui,gim} works have been 
done to explore electron transfer through different bridging molecular 
structures. A full quantum mechanical approach is needed~\cite{datta} 
to study electron transport in such molecular bridge systems where transport 
properties are characterized by several key factors like, quantization of 
energy levels, quantum interference of electronic waves associated with the 
geometry of bridging system adopts within the junctions and other several 
parameters of the Hamiltonian that are used to describe a complete system.

Here we use a simple parametric approach~\cite{sm1,sm2,sm3,sm4} rather 
than {\em ab initio} methods to describe conformation-dependent electron 
conductance in a biphenyl molecule. The physical picture about 
conformation-conductance correlation that emerges from our present study 
based on the single band TB model is exactly the same as obtained in the 
experiment~\cite{latha} and provides a very good insight to the problem.

\begin{figure}[ht]
{\centering \resizebox*{7.5cm}{2.2cm}{\includegraphics{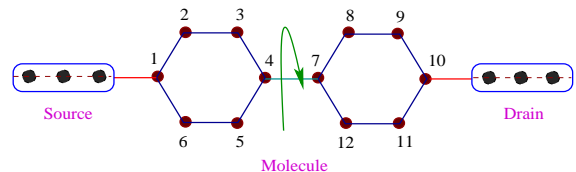}}\par}
\caption{(Color online). A biphenyl molecule attached to source and drain 
electrodes. The relative twist among the molecular rings is described by 
the green arrow.}
\label{biphenyl}
\end{figure}

Let us refer to Fig.~\ref{biphenyl} where a biphenyl molecule is connected 
to two semi-infinite one-dimensional ($1$D) non-interacting electrodes,
commonly known as source and drain. 
The single particle Hamiltonian for the entire system which describes 
the molecule and side-attached electrodes becomes,
\begin{equation}
H=H_{\mbox{\tiny M}} + H_{\mbox{\tiny ele}} + H_{\mbox{\tiny tun}}.
\label{equ1}
\end{equation}
The first term $H_{\mbox{\tiny M}}$ represents the Hamiltonian of the 
biphenyl molecule coupled to source and drain electrodes. Within a 
nearest-neighbor hopping approximation, the TB Hamiltonian of the 
molecule containing $12$ ($N=12$) atomic sites gets the form,
\begin{eqnarray}
H_{\mbox{\tiny M}} & = & \sum_i \epsilon c_i^{\dagger} c_i + \sum_i v 
\left[c_{i+1}^{\dagger} c_i + c_i^{\dagger} c_{i+1}\right] \nonumber \\
 & + & \sum_j \epsilon c_j^{\dagger} c_j + \sum_j v
\left[c_{j+1}^{\dagger} c_j + c_j^{\dagger} c_{j+1}\right] \nonumber \\
 & + & v_{4,7} \left[c_4^{\dagger}c_7 + c_7^{\dagger}c_4\right]
\label{equ2}
\end{eqnarray}
where the index $i$ is used for the left ring and for the right ring
we use the index $j$. $\epsilon$ represents the site energy of an electron 
at $i$-($j$-)th site and $v$ gives the nearest-neighbor coupling strength
between the molecular sites. $c_i^{\dagger}$($c_j^{\dagger}$) and 
$c_i$($c_j$) are the creation and annihilation operators, respectively, 
of an electron at the $i$-($j$-)th site. The last term in the right hand 
side of Eq.~\ref{equ2} illustrates the coupling among the molecular 
rings and in terms of the relative twist angle $\theta$ between these 
two rings, the coupling strength $v_{4,7}$ is written as 
$v_{4,7}=v \cos\theta$.

Similarly the second and third terms of Eq.~\ref{equ1} denote the TB 
Hamiltonians for the two semi-infinite $1$D electrodes and their couplings
to the molecule. They are expressed as follows.
\begin{eqnarray}
H_{\mbox{\tiny ele}} & = & H_{\mbox{\tiny S}} + H_{\mbox{\tiny D}} 
\nonumber \\
 & = & \sum_{\alpha={\mbox{\tiny S,D}}} \left\{\sum_n \epsilon_0 d_n^{\dagger} d_n 
+ \sum_n t_0 \left[d_{n+1}^{\dagger} d_n + h.c. \right]\right\}, 
\nonumber \\
\label{equ3}
\end{eqnarray}
and,
\begin{eqnarray}
H_{\mbox{\tiny tun}} & = & H_{\mbox{\tiny S,mol}} + H_{\mbox{\tiny D,mol}} 
\nonumber \\
& = & \tau_{\mbox{\tiny S}}[c_p^{\dag}d_0 + h.c.] + \tau_{\mbox{\tiny D}}
[c_q^{\dag}d_{N+1} + h.c.].
\label{equ4}
\end{eqnarray}
The parameters $\epsilon_0$ and $t_0$ correspond to the site energy and 
nearest-neighbor hopping integral in the source and drain electrodes. 
$d_{n}^{\dag}$ and $d_{n}$ are the creation and annihilation operators, 
respectively, of an electron at the site $n$ of the electrodes. The 
hopping integral between the source and the molecule is 
$\tau_{\mbox{\tiny S}}$, while it is  $\tau_{\mbox{\tiny D}}$ between the 
molecule and the drain. The source and drain are attached to the biphenyl 
molecule via the sites $p$ and $q$, respectively, those are variable.

To calculate two-terminal conductance ($g$) we use the Landauer conductance
formula $g=(2e^2/h)T$, where the transmission function
$T={\mbox{Tr}} \left[\Gamma_{\mbox{\tiny S}} \, G_{\mbox{\tiny M}}^r \,
\Gamma_{\mbox{\tiny D}} \, G_{\mbox{\tiny M}}^a\right]$~\cite{datta}.
Here, $G_{\mbox{\tiny M}}^r$ and $G_{\mbox{\tiny M}}^a$ are the retarded 
and advanced Green's functions, respectively, of the molecule including 
the effects of the electrodes.
$G_{\mbox{\tiny M}}=\left(E-H_{\mbox{\tiny M}}-\Sigma_{\mbox{\tiny S}}
-\Sigma_{\mbox{\tiny D}} \right)^{-1}$, where $\Sigma_{\mbox{\tiny S}}$ 
and $\Sigma_{\mbox{\tiny D}}$ are the self-energies due to coupling
of the chain to the source and drain, respectively, while 
$\Gamma_{\mbox{\tiny S}}$ and $\Gamma_{\mbox{\tiny D}}$ are their 
imaginary parts.

Throughout the analysis we choose the site energies in the molecule and 
side-attached electrodes to zero, $\epsilon=\epsilon_0=0$. The 
nearest-neighbor hopping integral in the electrodes ($t_0$) is set
at $2$eV, while in the molecule ($v$) it is fixed at $1$eV. The hopping
integrals of the molecule to the source and drain electrodes 
($\tau_{\mbox{\tiny S}}$ and $\tau_{\mbox{\tiny D}}$) are also set at 
$1$eV. Here, we consider that the entire voltage drop takes place across 
the molecule-electrode interfaces and it is a very good approximation for 
smaller size molecules. We also restrict ourselves at absolute zero 
temperature and choose the units where $c=e=h=1$. The energy scale 
is measured in unit of $v$.

\begin{figure}[ht]
{\centering \resizebox*{7.5cm}{7cm}{\includegraphics{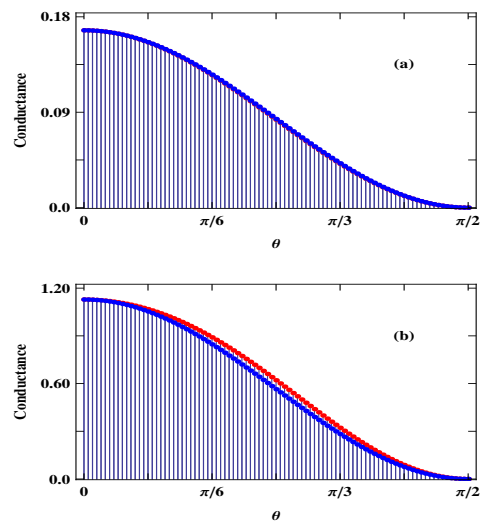}}\par}
\caption{(Color online). Electronic conductance for a specific energy as a 
function of twist angle for the biphenyl molecule when the electrodes are 
connected at the molecular sites $1$ ($p=1$) and $10$ ($q=10$), as shown in 
Fig.~\ref{biphenyl}. The results are computed for two typical energy values
where we choose $E=0.25$eV in (a) and in (b) we fix the energy $E$ at $1.65$eV.
The red dotted curve, drawn from numerical results, is superimposed on the 
blue dotted curve generated from the cosine squared relation: 
$A\cos^2(\theta)$, where $A$ represents the conductance amplitude at 
$\theta=0^\circ$.}
\label{conductance1}
\end{figure}

Figure~\ref{conductance1} describes the variation of electronic conductance
of the biphenyl molecule for a typical energy as a function of twist 
angle $\theta$ when the source and drain electrodes are attached to the
molecular sites $1$ and $10$, respectively. The results are shown for
two different energy values. In (a) we set $E=0.25$eV, while in (b) it 
is fixed at $1.65$eV. The red dotted curves in the spectra are generated 
from the numerical results and they are superimposed on the blue dotted
curves those are plotted from the cosine squared relation $A \cos^2(\theta)$,
\begin{figure}[ht]
{\centering \resizebox*{7.5cm}{7cm}{\includegraphics{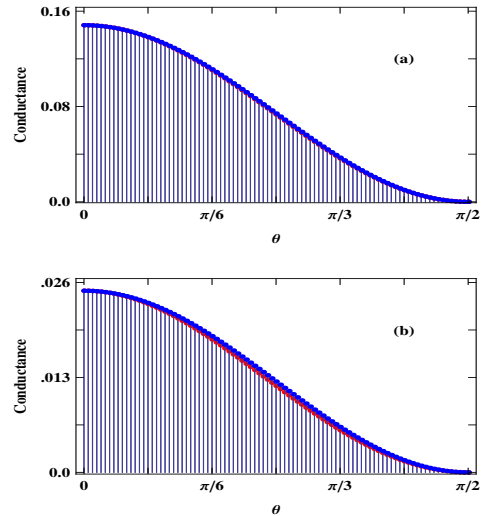}}\par}
\caption{(Color online). Same as Fig.~\ref{conductance1}, with $p=5$ and
$q=8$.}
\label{conductance2}
\end{figure}
where $A$ is the conductance amplitude for the planar conformation of the 
molecule. We evaluate this amplitude $A$ numerically. Very interestingly 
we notice that for $E=0.25$eV the red dotted curve sharply coincides with 
the blue one, and even for the other case i.e., when $E=1.65$eV the results 
are surprisingly close to each other. We also carry out extensive numerical 
work for other possible energies within the allowed energy band and find 
that the molecular conductance determined from the Landauer conductance 
formula agrees well with the cosine squared relationship. Thus we can 
emphasize that our numerical results can well fit the experimental data and 
provide a detailed information of the behavior of the molecular conductance
on its conformation. Now the reduction of electronic conductance with the 
molecular twist can be clearly understood from the following interpretation.
The degree of $\pi$-conjugation between the molecular rings decreases with
the rise of twist angle $\theta$ which results a reduction of the molecular 
conductance because the transfer rate of electrons through the biphenyl
molecule scales as the square of the $\pi$-overlap~\cite{nitz}. At the
typical case when $\theta$ reaches to $\pi/2$, the $\pi$-conjugation between
the molecular rings vanishes completely, and therefore, the conductance 
drops to zero. Obviously, it becomes a maximum for the planar conformation 
($\theta=0^{\circ}$) of the molecule. Thus, twisting one molecular ring 
with respect to the other electronic transmission through the biphenyl 
molecule may be controlled and eventually one can reach to the insulating 
phase. This phenomenon leads to a possibility of getting a switching action 
using this molecule.

Similar observations are presented in Fig.~\ref{conductance2} when the 
source and drain electrodes are coupled to the molecule at the sites 
$5$ ($p=5$) and $8$ ($q=8$), respectively. All the other parameters are
the same as in Fig.~\ref{conductance1}. It is interesting to note that the
quantum interference does not destroy the cosine squared dependence
between junction conductance and molecular conformation, which proves the 
robustness of the conformation-conductance correlation. Our numerical
results corroborate the experimental findings~\cite{latha}.

Before we end, it should be pointed out that though the results presented
in this communication are worked out for absolute zero temperature, they 
should be valid even for finite temperatures ($\sim 300\,$K) as the 
broadening of the energy levels of the biphenyl molecule due to its 
coupling with the electrodes will be much larger than that of the thermal 
broadening~\cite{datta,sm1,sm2,sm3,sm4}. Throughout our work, we 
numerically compute electronic conductance of the molecule for a typical 
set of parameter values and in our model calculations we choose them only 
for the sake of simplicity. Though the results presented here change 
numerically with these parameter values, but all the basic features remain 
exactly invariant which we confirm through our extensive numerical 
calculations.

The author is thankful to Prof. Abraham Nitzan for stimulating discussions.

\end{document}